\documentclass[pdftex]{PoS}
\usepackage{physics}
\usepackage{url}
\usepackage{comment}
\usepackage{simplewick}
\usepackage{cite}

\newcommand{\cev}[1]{\reflectbox{\ensuremath{\vec{\reflectbox{\ensuremath{#1}}}}}}
\newcommand{\Slash}[1]{{\ooalign{\hfil/\hfil\crcr\(#1\)}}}
\newcommand{\MSbar}{\overline{\mathrm{MS}}}

\title{Calculation of PCAC mass with Wilson fermion using gradient flow%
\\ \vspace*{-48mm}\hspace{4.7cm} \small{\textrm{UTHEP-745, UTCCS-P-130, J-PARC-TH-0212, KYUSHU-HET-204}} \vspace*{38mm}
}

\ShortTitle{Calculation of PCAC mass with Wilson fermion using gradient flow}

\author{\speaker{Atsushi Baba}, Asobu Suzuki\\
        Graduate School of Pure and Applied Sciences, University of Tsukuba, Tsukuba, Ibaraki,  305-8571, Japan\\
        E-mail: \email{ababa@het.ph.tsukuba.ac.jp}}
        
\author{Shinji Ejiri\\
        Department of Physics, Niigata University, Niigata 950-2181, Japan}
        
\author{Kazuyuki Kanaya\\
        Tomonaga Center for the History of the Universe, University of Tsukuba, Tsukuba, Ibaraki,  305-8571, Japan}
        
\author{Masakiyo Kitazawa\\
        Department of Physics, Osaka University, Osaka 560-0043, Japan\\
        J-PARC Branch, KEK Theory Center, Institute of Particle and Nuclear Studies, KEK, 203-1, Shirakata, Tokai, Ibaraki, 319-1106, Japan} 

\author{Hiroshi Suzuki\\
        Department of Physics, Kyushu University, 744 Motooka, Nishi-ku, Fukuoka 819-0395, Japan}

\author{Yusuke Taniguchi\\
        Center for Computational Science (CCS), University of Tsukuba, Tsukuba, Ibaraki,  305-8571, Japan}

\author{Takashi Umeda\\
        Graduate School of Education, Hiroshima University, Higashihiroshima, Hiroshima 739-8524, Japan}

\abstract{
We calculate the PCAC mass for $(2+1)$ flavor full QCD with Wilson-type quarks.
We adopt the \textbf{S}mall \textbf{F}low-\textbf{\textit{t}}ime e\textbf{X}pansion (\textbf{SF\textit{t}X}) method based on the gradient flow which provides us a general way to compute correctly renormalized observables even if the relevant symmetries for the observable are broken explicitly due to the lattice regularization, such as the Poinc\'{a}re and chiral symmetries.
Our calculation is performed on heavy $u, d$ quarks mass ($m_{\pi}/m_{\rho}\simeq0.63$) and approximately physical $s$ quark mass with fine lattice $a \simeq 0.07$~fm.
The results are compared with those computed with the Schr\"odinger functional method.
}

\FullConference{37th International Symposium on Lattice Field Theory - Lattice2019\\
		16-22 June 2019\\
		Wuhan, China}

\begin{document}
\section{Introduction} \label{sec:intro}
Though the quark masses are fundamental parameters of QCD,
they cannot be measured experimentally since quarks are confined in hadrons.
Here, non-perturbative calculation by lattice QCD plays an important role to determine the quark masses.
When we calculate the renormalized quark mass, the PCAC mass is often used.
The PCAC mass is the quark mass parameter appearing in the PCAC relation, which is a chiral Ward-Takahashi identity given by
\begin{align}
	\left< 0 \right| \left\{ \partial_{\mu} A^{a}_{\mu} (x) + 2m_{f} P^{\,a}(x) \right\} \mathcal{O}(y) \left |0 \right> + \left< 0\right| \delta^{a}_{x} \mathcal{O}(y) \left|0 \right>  = 0,
\end{align}
where $m_{f}$ is the PCAC mass and $\delta^{a}_{x}$ means the infinitesimal chiral transformation.
Axial-vector current $A^{a}_{\mu}(x)$ and pseudo-scalar density $P^{\,a}(x)$ are defined by
\begin{gather}
	A^{a}_{\mu}(x) = \bar{\psi}_{f}(x) \gamma_{5} \gamma_{\mu} T^{a} \psi_{f}(x) ,
	\\
	P^{\,a}(x) = \bar{\psi}_{f}(x) \gamma_{5} T^{a} \psi_{f}(x),
\end{gather}
and $\mathcal{O}(x)$ is an operator which we can set arbitrarily.
When we set $\mathcal{O}(y) = P^{\,a}(y)$ and integrate over the spacial coordinates, we obtain a PCAC relation
\begin{align}
	\left< \partial_{0} \, A^{a}_{\mu} (x_0) P^{\,a}(0) \right> = -2m_{f} \left< P^{\,a}(x_0) P^{\,a}(0) \right>,
\end{align}
with which we can calculate the PCAC mass by
\begin{align}
	m_{f} =- \frac{\left< \partial_{0} \, A^{a}_{0} (x_{0}) P^{\,a}(0) \right>}{2 \left< P^{\,a}(x_0) P^{\,a}(0) \right>}.
\end{align}
 
Recently, a new use of the gradient flow method \cite{Narayanan:2006rf, GF1, GF2, GF3} was proposed to calculate correctly renormalized observables \cite{suzuki1,suzuki2}.
The new method is called \textbf{Small Flow-\textit{t}ime eXpansion (SF\textit{t}X) method}.
Making use of the finiteness of flowed operators, non-perturbative estimates of observables are extracted by taking a vanishing flow-time extrapolation.
The SF\textit{t}X method was first applied to evaluate the energy-momentum tensor for which the explicit violation of the Poincar\'e invariance on the lattice has been a hard obstacle in obtaining a non-perturbative estimate \cite{EMT1,EMT2}.

Because the SF\textit{t}X method is generally applicable to any observables including chiral observables~\cite{suzuki2}, we are applying it to QCD with dynamical quarks~\cite{EMT2,TS2}.
In this paper, we study the PCAC mass by the SF\textit{t}X method in QCD with $(2+1)$-flavors of improved Wilson quarks.

\section{SF\textit{t}X method} \label{sec:sftx}
In this study, we adopt the simplest gradient flow for the gauge field \cite{GF1}: 
\begin{align} \label{eq:GF_gauge}
	\partial_{t} B_{\mu} (t, x) = D_{\nu} G_{\nu \mu} (t, x) ,\qquad B_{\mu} (0, x) = A_{\mu} (x),
\end{align}
where the field strength $G_{\nu \mu}$ and the covariant derivative $D_{\nu}$ are defined in terms of the flowed gauge field $B_{\mu}$.
The flow equations for quarks are given by~\cite{GF3}: 
\begin{gather}\label{eq:GF_fermion}
	\partial_{t} \chi_f (t, x) = D^2 \chi_f (t, x) ,\qquad \chi_f (0, x) = \psi_f(x) ,
	\\
	\partial_{t} \bar{\chi}_f (t, x) = \bar{\chi}_f (t, x) \cev{D} \,^2 ,\qquad \bar{\chi}_f (0, x) = \bar{\psi}_f (x) ,
\end{gather}
with $D_{\mu} \chi_f (t, x) = \left( \partial_{\mu} + B_{\mu} (t, x) \right) \chi_f$ and
$\bar{\chi}_f (t, x) \cev{D}_\mu = \bar{\chi}_f (t, x) \left( \cev{\partial_{\mu}} - B_{\mu} (t, x) \right)$.

In terms of the flowed fields, the correctly renormalized axial-vector current and pseudo-scalar density in the $\MSbar$ scheme at $\mu = 2$ GeV is given by~\cite{suzuki2}: 
\begin{gather}
A^{a}_{\mu} (x) = \lim _{t \to 0} A^{a}_{\mu} (t, x) = \lim _{t \to 0} \, c_A (t)  \varphi_f (t) \, \bar{\chi}_f(t, x) \gamma_{5} \gamma_{\mu} T^{a} \chi_f(t, x),
\\
P^{\,a} (x) = \lim_{t\ \to 0} P^{\,a} (t, x) = \lim _{t \to 0} \, c_S (t)  \varphi_f (t) \, \bar{\chi}_f(t, x) \gamma_{5} T^{a} \chi_f(t, x),
\end{gather}
where the matching coefficients $c_A(t), c_S(t)$ and fermion wave function renormalization factor $\varphi (t)$ are
\begin{gather}
	c_A (t) = \left\{ 1 + \frac{\bar{g}^2( \mu^{\prime} )}{(4 \pi)^2} \left[ - \frac{3}{2} + \frac{4}{3} \ln 432 \right] \right\} ,
	\\
	c_S (t) = \left\{ 1 + \frac{\bar{g}^2( \mu^{\prime} )}{(4 \pi)^2} \left[ 4 \left( \ln (2t\mu^{\prime2}) + \gamma_E \right) + 8 + \frac{4}{3} \ln 432 \right] \right\} \frac{\bar{m}_f (\mu^{\prime})}{\bar{m}_f(2 \mathrm{GeV})}, 
	\\
	\varphi_f (t) = \frac{-6}{ ( 4\pi)^2 t^2 \expval{ \bar{\chi}(t, x) \overset{\leftrightarrow}{\Slash{D}} \chi(t, x) } } ,
\end{gather}
where $\bar{g}(\mu^{\prime})$ and $\bar{m}(\mu^{\prime})$ are running coupling and running mass in the $\MSbar$ scheme at the renormalization scale $\mu^{\prime}(t)$, and $\gamma_E$ is the Euler-Mascheroni constant.
Then, the PCAC mass is given by 
\begin{align} \label{eq:PCAC_mass}
	m_{f} = \lim_{t \to 0}\; m_{f}(t) = - \lim_{t \to 0}\, \frac{ c_A(t) c_S(t) \varphi_{f}^{2} (t) \left< \partial_{0} \, A^{a}_{0} (t, x_{0}) P^{\,a}(t, 0) \right>}{2 c_S^{2}(t) \varphi_{f}^{2}(t) \left< P^{\,a}(t, x_0) P^{\,a}(t, 0) \right>}.
\end{align}

Final results of $m_f$ should be independent of the scale $\mu^{\prime}(t)$ as far as it is $O(1/ \sqrt{t})$ to preserve the quality of the perturbation theory. 
A conventional choice is $\mu^{\prime} = \mu_{d}(t) \equiv 1/\sqrt{8t}$, which is a natural scale of flowed observables because the gradient flow smears the fields over a physical extent of $\sim \sqrt{8t}$ \cite{GF1}.
Recently, a new choice was proposed by Harlander \textit{et al.}\ as $\mu^{\prime} = \mu_{0}(t) \equiv 1 / \sqrt{2 e^{\gamma_E} t}$ \cite{HKL}.
Because $\mu_{0} \simeq 1.5 \mu_{d}$ , we expect that, in asymptotically free theories, the range of $t$ in which the perturbative expansion is well applicable is extended towards larger $t$ with the $\mu_{0}$-scale than the $\mu_d$-scale.
In a study of Ref.~\cite{physicalpoint} on the energy momentum tensor and chiral observables in finite-temperature QCD, we found that the wider range of $t$ with the $\mu_{0}$-scale is helpful in reducing systematic uncertainties from the $t\to0$ extrapolation.
We test the $\mu_0$-scale also in this study.

We evaluate Eq.~(\ref{eq:PCAC_mass}) non-perturbatively by performing lattice simulations.
The original procedure of the SF$t$X method is to take the continuum limit $a \to 0$ first, then the leading small-$t$ correction to the flowed PCAC mass will be $m_{f} (t) = m_{f} + t A + O(t^{2})$, where $A$ is the contamination from dimension-five operators.
In Refs.~\cite{EMT2,TS2}, an alternative procedure to take $t \to 0$ limit before the continuum limit was proposed. 
To the leading order of $O(a^2)$ we will have additional contaminations like 
\begin{align}\label{eq:lattice_artifact}
m_{f} (t, a) = m_{f} (t) + O( a^{2}\! / t,\; a^{2} T^{2}\!,\; a^{2} m^{2}\!,\; a^{2} \Lambda_{\mathrm{QCD}}^{2}).
\end{align}
Among the $O(a^2)$ terms, the term $O(a^{2}\! / t)$ is singular in the $t \to 0$ extrapolation.
We may avoid this difficulty by identifying a range of $t$, ``linear window'', a range of $t$ in which terms like $O(a^{2}\! / t)$ and $O(t^2)$ are not dominating, and taking a $t\to0$ extrapolation using the data in the linear window.
We may then evaluate the RHS of Eq.~(\ref{eq:PCAC_mass}) by succeeding $a\to0$ extrapolation to remove the remaining $O(a^{2} T^{2},\; a^{2} m^{2},\; a^{2} \Lambda_{\mathrm{QCD}}^{2})$ lattice artifacts. 
We may check the validity of the linear windows by performing non-linear fits including $O(a^{2}\! / t)$ and $O(t^2)$ terms.
The difference between the linear and non-linear fits gives an estimate of the systematic error due to the fit ansatz.
See Ref.~\cite{EMT2} for more details.

\section{Lattice Setup} \label{sec:setup} 

We study $(2 + 1)$-flavor QCD adopting a non-perturbatively $O(a)$-improved Wilson quark action and the RG-improved Iwasaki gauge action.
We choose a set of CP-PACS+JLQCD configurations generated at $\beta = 2.05$ corresponding to $a \simeq 0.07$ fm, degenerate $u$, $d$ quark mass corresponding to $m_{\pi}/m_{\rho} \simeq 0.63$, and almost physical $s$ quark mass corresponding to $m_{\eta_{ss}}/m_{\phi} \simeq 0.74$ on a $28^3\times56$ lattice~\cite{zero_temp_conf}.
At this simulation point, the $O(a)$-improvement of axial-vector current $c_{A}$ is culculated by CP-PACS/JLQCD and ALPHA Collaborations as $c_{A} = -0.0272(18)$~\cite{axial_imp}.
We employ the 5-loop order $\beta$-function~\cite{5loop_beta_function} and anomalous dimension~\cite{5loop_anomalous_dimension} to calculate running coupling and running mass  in matching coefficients $c_A(t)$ and $c_S(t)$.

In the study of Ref.~\cite{zero_temp_conf }, PCAC masses at each simulation points have been calculated by the Schr\"odinger functional method. 
The bare PCAC quark masses at the simulation point of this study using the same configurations are $am_{ud} = 0.02105(17)$ and $am_{s} = 0.03524(26)$, which correspond to
\begin{align} \label{eq:PCACmass_SF}
	m^{SF}_{u} = 82.3 \pm 4.1, \qquad m^{SF}_{s} = 137.9 \pm 6.8,
\end{align}
in MeV unit.

\section{Numerical results} \label{sec:results}
\begin{figure}[tb]
\begin{minipage}{0.5\hsize}
	\centering
	\includegraphics[width=\linewidth]{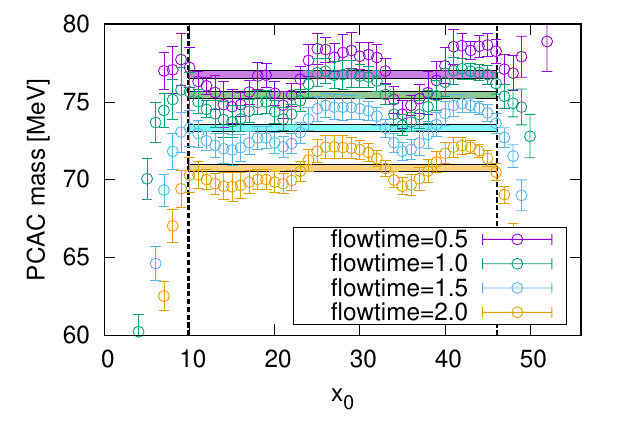}
\end{minipage}
\begin{minipage}{0.5\hsize}
	\centering
	\includegraphics[width=\linewidth]{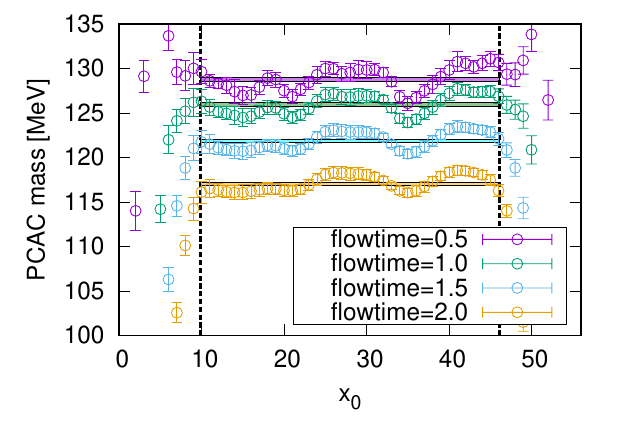}
\end{minipage}
\caption{
PCAC mass of $u$ quark (\textbf{left}) and $s$ quark (\textbf{right}) as function of the Euclidean time $x_0$, at flow-time $t/a^{2} = 0.5$ (violet), $1.0$ (green), $1.5$( cyan) and $2.0$ (yerrow). 
The $\mu_0$-scale was adopted. 
Vertical lines indicate the range of constant fit.
Fit range is the same for all flow-times.
Errors are statistical only, estimated by the jackknife method.
}
\label{fig:PCAC_u_s}
\end{figure}

\begin{figure}[tb]
\begin{minipage}{0.5\hsize}
	\centering
	\includegraphics[width=\linewidth]{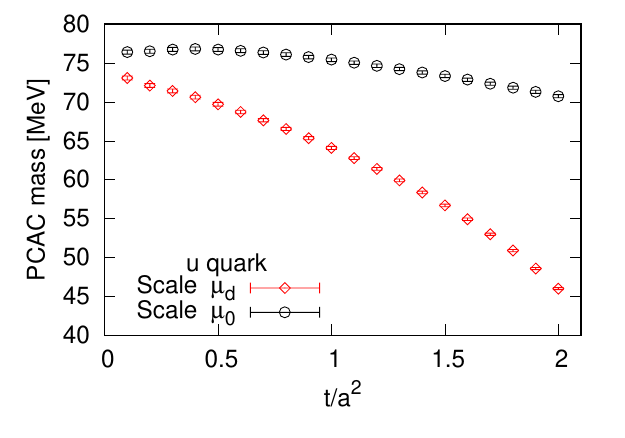}
\end{minipage}
\begin{minipage}{0.5\hsize}
	\centering
	\includegraphics[width=\linewidth]{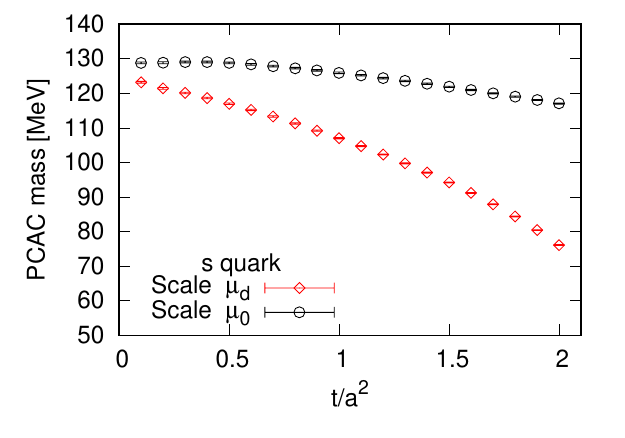}
\end{minipage}
\caption{%
PCAC mass of $u$ quark (\textbf{left}) and $s$ quark (\textbf{right}) as function of the flow-time time $t/a^{2}$.
Red diamonds and black circles are the results of $\mu_{d}$- and $\mu_{0}$-scales, respectively.
}
\label{fig:PCAC_constfit}
\end{figure}

In Fig.~\ref{fig:PCAC_u_s}, we show the PCAC mass for $u$ and $d$ quarks computed with the $\mu_{0}$-scale as function of Euclidean time at four different flow-times.
We perform constant fits at each flow-time within the range indicated by the vertical lines.
We use the same fit range for all flow-times.
The results of the fits are shown by colored bands.
The errors of the constant fits are statistical only, estimated using the jackknife method.

In Fig.~\ref{fig:PCAC_constfit}, we show the PCAC mass as function of flow-time.
Red diamonds and black circles are the results with $\mu_{d}$- and $\mu_{0}$-scales, respectively.
We see that, with the conventional $\mu_{d}$-scale, it is not well unambiguous to identify a linear window due to the bend at large $t$, and thus the $t \to 0$ extrapolation is sensitive to the choice of linear window.
On the other hand, with the $\mu_{0}$-scale, we see a linear behavior in a wider range of $t$, which enables us to carry out a much more stable and reliable $t\to0$ extrapolation of Eq.~(\ref{eq:PCAC_mass}).
We thus adopt the $\mu_0$-scale to calculate the PCAC masses.

\begin{figure}[tb]
\begin{minipage}{0.5\hsize}
	\centering
	\includegraphics[width=\linewidth]{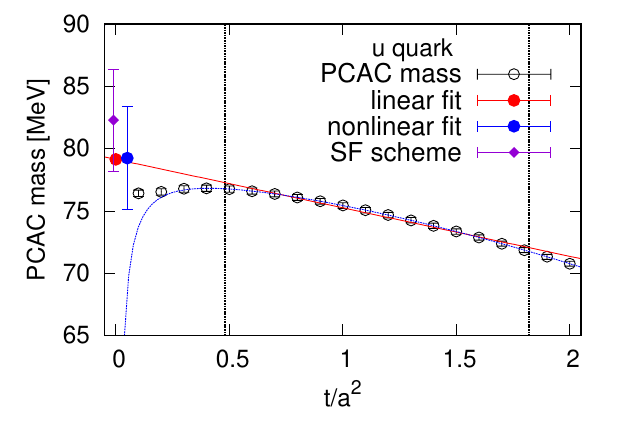}
\end{minipage}
\begin{minipage}{0.5\hsize}
	\centering
	\includegraphics[width=\linewidth]{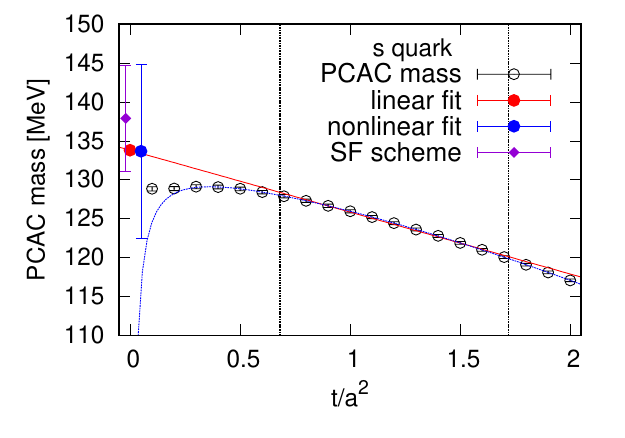}
\end{minipage}
\caption{%
PCAC mass of $u$ quark (\textbf{left}) and $s$ quark (\textbf{right}) as function of flow-time.
The $\mu_0$-scale is adopted. 
Vertical lines indicate the linear window we adopt.
We take $t \to 0$ extrapolation by a linear fit (red) and estimate the systematic error by the difference from the result of non-linear fit (blue).
The violet diamond at $t/a^2\sim0$ is the PCAC mass evaluated by the Schr\"odinger functional method~\cite{zero_temp_conf}.
}
\label{fig:nonlinear_fit}
\end{figure}

The results of PCAC masses with $\mu_{0}$-scale are summarized in Fig.~\ref{fig:nonlinear_fit} as function of flow-time $t/a^{2}$ for the $u$ quark (left panel) and for the $s$ quark (right panel).
We identify linear windows as follows:
First of all, we require the flow-time to satisfy 
$
	a \le \sqrt{8t} \le \min(N_t a/2,N_s a/2),
$
\emph{i.e.}, the smearing range $\sqrt{8t}$ by the gradient flow should be larger than the minimal lattice sepatration to make the smearing effective,
and smaller than the half of the smallest lattice extent to avoid finite-size effects due to overlapped smearing.
We then look for a range of $t$ in which terms linear in $t$ look dominating, and try linear extrapolation with various choices of the fitting range. 
We then select a (temporally) best linear fit whose fitting range is the widest under the condition that $\chi^2/N_\textrm{dof}$ is smaller than a cutoff value.
In this study, due to limitation of the statistics, we disregard correlations among data at different $t$. 
Thus the absolute value of $\chi^2/N_\textrm{dof}$ does not have a strong sense --- we vary the cutoff value widely.
In this study, consulting the stability of the fit results, we choose 1.0 as the cutoff value for PCAC mass.
The linear window we adopt is shown by the two vertical lines in~Fig.~\ref{fig:nonlinear_fit}.

To confirm the validity of the linear window and to estimate a systematic error due to the fit ansatz, we also make 
additional non-linear fit of the form $m_{f} (t, a) = m_{f} + t A + t^2 B + \frac{a^2}{t} C$,
using the data within the linear window.
Results of linear and non-linear fits are shown by red and blue lines in~Fig.~\ref{fig:nonlinear_fit}.
We adopt the results of the linear fits as central values and take the difference between the two fits as an estimate of the systematic error due to the fit ansatz.
Our results of the PCAC masses are
\begin{align} \label{eq:PCACmass_GF}
	m^{\mathrm{SF}t\mathrm{X}}_{u} = 79.14 \pm 0.19, \qquad m^{\mathrm{SF}t\mathrm{X}}_{s} = 133.81 \pm 0.24,
\end{align}
in MeV unit, where statistical error and systematic error due to fit ansatz are included.

\section{Summary and outlook} \label{sec:summary}

We studied PCAC mass in lattice QCD with $(2 + 1)$-flavors of dynamical Wilson quarks.
Nonperturbative renormalization is carried out by the SF\textit{t}X method based on the gradient flow.
Our calculation was performed at heavy $u,$ $d$ quarks mass ($m_{\pi}/m_{\rho}\simeq0.63$) and approximately physical $s$ quark mass on a fine lattice with $a \simeq 0.07$ fm.
As the renormalization scale in the SF\textit{t}X method, we adopt the recently proposed $\mu_{0}$-scale.
We found that the $\mu_{0}$-scale is helpful to reduce uncertainty in the $t\to0$ extrapolation.

Our results for the PCAC masses for $u$ (or $d$) quark and $s$ quark are given in Eq.~(\ref{eq:PCACmass_GF}).
These are consistent with the results of conventional Schr\"odinger functional method, Eq.~(\ref{eq:PCACmass_SF}), obtained on the same configurations.
By virtue of the gradient flow, statistical error is well suppressed compered with the results of the Schr\"odinger functional method.

We are extending the study to $(2+1)$-flavor QCD with physical $u$, $d$ and $s$ quarks~\cite{physicalpoint}.
To obtain final results, we also have to repeat the calculation at different lattice spacings to carry out the continuum extrapolation.

\acknowledgments 
This work was in part supported by JSPS KAKENHI Grant Numbers 
JP19K03819, JP19H05146, JP19H05598, JP18K03607, JP17K05442 and JP16H03982.
This research used computational resources of COMA, Oakforest-PACS, and Cygnus provided by the Interdisciplinary Computational Science Program of Center for Computational Sciences, University of Tsukuba,
K and other computers of JHPCN through the HPCI System Research Projects (Project ID:hp17208, hp190028, hp190036) and JHPCN projects (jh190003, jh190063), OCTOPUS at Cybermedia Center, Osaka University, and ITO at R.I.I.T., Kyushu University.
The simulations were in part based on the lattice QCD code set Bridge++ \cite{bridge}.

\end{document}